\documentstyle[aps]{revtex}
\begin{document}
\draft

\title{On a network model of localization in a random magnetic field}

\author{Yong Baek Kim,\cite{YBK} Akira Furusaki,\cite{AF} and Derek K. K. Lee}
\address{Department of Physics, Massachusetts Institute of
Technology, Cambridge, Massachusetts 02139}
\date{\today}
\maketitle

\begin{abstract}
We consider a network model of snake states to study the
localization problem of non-interacting fermions
in a random magnetic field with zero average.
After averaging over the randomness, the network of snake
states is mapped onto $M$ coupled
SU$(2N)$ spin chains in the $N \rightarrow 0$ limit.
The number of snake states near the zero-field contour, $M$,
is an even integer.
In the large conductance limit $g = M {e^2 \over 2 \pi \hbar}$
($M \gg 2$), it turns out that this system is
equivalent to a particular representation of the
${\rm U}(2N) / {\rm U}(N) \times {\rm U}(N)$ sigma model ($N \rightarrow 0$)
{\it without} a topological term.
The beta function $\beta (1/M)$ of this sigma model in the
$1/M$ expansion is consistent with the previously known
$\beta (g)$ of the unitary ensemble.
These results and further plausible arguments
support the conclusion that all the states
are localized.

\end{abstract}

There has been much recent interest in the quantum motion of a charged
particle in a random perpendicular magnetic field in two dimensions.
Theoretically, this problem arises from the study of gauge theories of
strongly-correlated electronic systems. An example is the Chern-Simons
theory for the half-filled Landau level\cite{hlr,kz}. In a mean-field
treatment, one obtains an effective theory of free fermions moving in
a weak effective magnetic field due to the overall cancellation of the
external field by the Chern-Simons field. This effective field has
zero average but contains a random static component induced by the
inhomogeneous electron density. Another example is the gauge theory of
the doped Mott insulators\cite{ioffe,nagaosa}. In this case, slow
fluctuations of the gauge field can be approximated as a static random
field.

Single-particle motion in a random magnetic field with zero average is
also interesting in the context of localization theory. One might
expect that a system with random magnetic fields should belong to the
same universality class as other disordered systems with broken
time-reversal symmetry and zero Hall conductance.  This is the unitary
class for which all states are localized, according to conventional
scaling theory\cite{palee}. In field-theoretic terms, the system would
be described at large distances by a non-linear sigma model (without a
topological term).  However, Zhang and Arovas\cite{zhang} have
recently suggested that long-range interactions between topological
densities in the non-linear sigma model, induced by local fluctuations
of $\sigma_{xy}$, may give rise to a novel delocalization transition
when the conductance of the system at short distances exceeds a
critical value of the order of $e^2/h$.

In this paper, we will present an analytic approach to this random
flux problem in the limit of smooth disorder, based on the random
network model developed in Ref. \cite{derek2} and the method
introduced by D.H. Lee and coworkers \cite{wang,dhlee,dhlee2} for the
discussion of the Chalker-Coddington network model\cite{chalker} in
the context of the quantum Hall transitions. We find that the
transport properties of the random-magnetic-field problem can be described in
the large-conductance limit by a non-linear sigma model without a
topological term.
Therefore, we argue that the localization properties of this system
are the same as that of the conventional unitary ensemble, suggesting
that all states are localized.
Similar conclusions have also been reached for the case
of weak and short-ranged disorder\cite{aronov,falko}.

There have been several numerical studies of this problem with
conflicting conclusions. The difficulty in the interpretation of the
numerical data arises from the fact that the localization length
increases rapidly as a function of energy as one approaches the band
center. Some authors\cite{avishai,kwaz,liu,sheng} have argued that
there exists a range of energies for which the eigenstates are
extended, while others\cite{derek1,derek2,sugi} suggest that all the
states are localized.  Sugiyama and Nagaosa\cite{sugi} have observed
that the numerical data are consistent with the conventional
single-parameter scaling of the conductance in the unitary
universality class. Although their numerical data are restricted to
bare conductances below $0.2 e^2/h$, we will argue that the
conventional scaling function continues to apply in the
large-conductance limit of this problem, in contrast with the
Zhang-Arovas theory.

We consider in this paper the case of a random magnetic field with
zero average $\langle B ({\bf r}) \rangle = 0$ and a non-zero variance
$\langle B^2 ({\bf r}) \rangle = B^2_0$.  We will work in the
semiclassical limit in which the correlation length $\lambda$ of the
random magnetic field is large compared to the typical magnetic length
$l_0 = (\hbar c / e B_0)^{1/2}$. In this limit of smooth disorder, an
electron moves along contours of constant magnetic field in a
direction determined by the sign of the field gradient on the
contour. This chirality reflects the broken time-reversal symmetry of
the problem. Since we are interested in the possibility of
delocalization, we can focus on contours which percolate geometrically
across the system. These are the $B=0$ contours where the classical
motion of the electron describe snake-like trajectories\cite{muller}.
To simplify further, we represent these zero-field contours as links
on a square network, carrying a fixed number, $M$, of the electronic
``snake states''. The nodes of the network represent the saddle points
of the magnetic field, where a particle is scattered to the right or
left.  This scattering, which is chaotic even in the classical limit,
is modeled by random amplitudes chosen from a distribution which
ensures that the network as a whole has zero Hall conductance. Quantum
interference due to these scattering processes is modeled by random
phase shifts and the mixing of the snake states along the links.

Details of the mapping of the semiclassical limit to the network model
can be found in Ref.~\cite{derek2}. We will emphasize here only the
points of particular relevance to the present discussion. It should be
noted that the electron states on the zero-field contours typically
occur {\it in pairs} at a given energy. This can be seen most easily
by considering the special case of a straight contour, across which
the field switches abruptly between two values, $\pm B_0$\cite{mitya}.
Restricting attention to the energies between the lowest two Landau
levels in the bulk, it can be shown that two modes propagate along the
contour in the same direction, arising ultimately from symmetric and
antisymmetric linear combinations of the Landau levels on either side
of the contour. Similarly, at a higher energy, each bulk Landau level
below that energy is associated with two states on the zero-field
contour. Thus, each link of the network carries an {\em even} number,
$M$, of channels propagating in the same direction.

The conductance of the system at short distances is given
by \cite{mitya}
\begin{equation}
G = M {e^2 \over 2 \pi \hbar} \ .
\label{conductance}
\end{equation}
A schematic representation of the network is given in Fig.~1 for the
two-channel case which was studied numerically in Ref.~\cite{derek1}.
In this paper, we are interested in the limit of large conductance so
that $ M \gg 2$. It should be pointed out that the semiclassical
picture of motion confined to contours of constant field breaks down
when the energy of the particle is so high that the lateral extent of
the wavefunction becomes comparable to the correlation length of the
magnetic field.  In terms of the number of channels, this condition
for the validity of the semiclassical picture is given by $M < M_{\rm
max} = (E /E_{\rm slope})^{3/4}$, where $E_{\rm slope} = {e B_0 \over
mc} (l_0/\lambda)^{2/3}$\cite{derek2}. Nevertheless, we can obtain an
arbitrary $M_{\rm max}$ by choosing a sufficiently large correlation
length.

Following Ref.~\cite{dhlee}, we represent the network model using the
schematic picture depicted in Fig.~2 for the two-channel case. Let us
take each zigzag line as a constant-$x$ curve and measure the
displacement along it by the coordinate $y$. We use the coordinate
system where the saddle points are located at $(x, y) = (n_x, n_y)$,
where $n_x$ and $n_y$ are integers. Taking the continuum limit in the
$y$-direction, the network can be represented by the
Hamiltonian\cite{wang,dhlee,dhlee2}
\begin{eqnarray}
H &=& \sum_x (-1)^x \int dy \sum_j \psi^{\dagger}_{j} (x,y)
{\partial_y \over i} \psi_j (x,y) \cr
&&- \sum_x \int dy \sum_j
\left [ t_j (x,y) \ \psi^{\dagger}_j (x+1,y) \psi_j (x,y) +
{\rm h.c.} \right ] \cr
&&- \sum_x \int dy \sum_{j < k}
\left [ t_{jk} (x,y) \ \psi^{\dagger}_j (x,y) \psi_k (x,y) +
{\rm h.c.} \right ] \ ,
\label{originalH}
\end{eqnarray}
where $\psi_j$ ($j = 1, \cdots, M$) is the fermion annihilation
operator of the $j$th snake state. The scattering processes of the
network are now modeled by tunneling ($t_j$) between neighboring sites with
the same channel index, as well as tunneling ($t_{jk}$) between
different channels at the same $x$-site.
We choose $t_j(x,y)$ and $t_{jk}(x,y)$ to be independent random complex
numbers with Gaussian distributions. The random magnitudes of these
matrix elements incorporate local fluctuations of Hall conductance
explicitly. In addition, the random phases will give rise to
interference effects which may lead to the localization of these
states. The two-channel network studied in Refs. \cite{derek1,derek2}
correspond here to the case where $|t_1|$ and $|t_2|$ are chosen such
that there is no bias for a particle to be scattered to the right or
left at each node (if the node is considered in isolation).
Since the effects of changing the energy, $E$, can be absorbed into
a dependence of the tunneling matrix elements, $t_j(x,y)$ and
$t_{jk}(x,y)$, on $E$, we can set $E = 0$ without loss of generality.
Note that, as we change $E$,
both of the tunneling matrix elements and the number of
snake states, $M$, change as functions of $E$.

In order to study transport properties, we have to consider the generating
functional for the retarded and advanced Green's
functions:\cite{stone,pruisken}
\[
Z = \int D {\bar \psi} D \psi \ e^{-S} \ ,
\]
where
\begin{eqnarray}
S &=& \sum_x \int dy \sum_j \sum_p i \eta S_p \
{\bar \psi}_{pj} (x,y) \psi_{pj} (x,y) -
H (\psi_j \rightarrow \psi_{pj},
\psi^{\dagger}_j \rightarrow {\bar \psi}_{pj}) \cr
&=& \sum_x \int dy \sum_j \sum_p \left [ i \eta S_p \
{\bar \psi}_{pj} (x,y) \psi_{pj} (x,y) -
(-1)^x {\bar \psi}_{pj} (x,y)
{\partial_y \over i} \psi_{pj} (x,y) \right ] \cr
&&+ \sum_x \int dy \sum_j \sum_p
\left [ t_j (x,y) \ {\bar \psi}_{pj} (x+1,y) \psi_{pj} (x,y) +
{\rm h.c.} \right ] \cr
&&+ \sum_x \int dy \sum_{j < k} \sum_p
\left [ t_{jk} (x,y) \ {\bar \psi}_{pj} (x,y) \psi_{pk} (x,y) +
{\rm h.c.} \right ] \ .
\label{originalS}
\end{eqnarray}
Here, $\psi_{pj}$ and ${\bar \psi}_{pj}$ are Grassmann variables.
Also, $p = + (-)$ represents the retarded (advanced) sector, $S_p =
{\rm sgn} (p)$, and $\eta$ is a positive infinitesimal number.

Now we let $\psi_{pj} \rightarrow \psi_{pj} (i\psi_{pj})$ and
${\bar \psi}_{pj} \rightarrow -i{\bar \psi}_{pj} ({\bar \psi}_{pj})$
for even (odd) $x$'s\cite{dhlee}.
Then, the action becomes
\begin{eqnarray}
S &=& \sum_x \int dy \sum_j \sum_p \left [ (-1)^x \eta S_p \
{\bar \psi}_{pj} (x,y) \psi_{pj} (x,y) +
{\bar \psi}_{pj} (x,y) \partial_y \psi_{pj} (x,y) \right ] \cr
&&+ \sum_x \int dy \sum_j \sum_p
\left [ t_j (x,y) \ {\bar \psi}_{pj} (x+1,y)
\psi_{pj} (x,y) + {\rm h.c.} \right ] \cr
&&- i \sum_x \int dy \sum_{j < k} \sum_p (-1)^x
\left [ t_{jk} (x,y) \ {\bar \psi}_{pj} (x,y)
\psi_{pk} (x,y) +
{\rm h.c.} \right ] \ .
\label{transformS}
\end{eqnarray}
Next we replicate the action and average over $t_j$ and $t_{jk}$ to get
\begin{eqnarray}
S &=& \sum_x \int dy \sum_j \sum_{\alpha}
[ (-1)^x \eta S_p \ {\bar \psi}_{\alpha j} (x,y) \psi_{\alpha j} (x,y) +
{\bar \psi}_{\alpha j} (x,y) \partial_y \psi_{\alpha j} (x,y) ] \cr
&&- \sum_x \int dy \sum_j \sum_{\alpha, \beta} \
\langle |t_j|^2 \rangle \
{\bar \psi}_{\alpha j} (x+1,y) \psi_{\beta j} (x+1,y)
{\bar \psi}_{\beta j} (x,y) \psi_{\alpha j} (x,y) \cr
&&+ \sum_x \int dy \sum_{j < k} \sum_{\alpha, \beta} \
\langle |t_{jk}|^2 \rangle \
{\bar \psi}_{\alpha j} (x,y) \psi_{\beta j} (x,y)
{\bar \psi}_{\beta k} (x,y) \psi_{\alpha k} (x,y) \ ,
\label{replicaS}
\end{eqnarray}
where $\alpha = (p,n)$ and $n = 1, \cdots, N$ is the replica index so that
$\alpha$ and $\beta$ take $2N$ values.

Taking $y$ as the imaginary time, we can see that the above action
describes a many-body system of interacting fermions on $M$
one-dimensional chains. We can further substitute the fermion
operators with the generators of SU$(2N)$:
\[
S^\alpha_{j, \beta} \equiv \psi^{\dagger}_{\alpha j}
\psi_{\beta j} - \delta_{\alpha \beta}
{1 \over 2N} \sum_\gamma \psi^{\dagger}_{\gamma j}
\psi_{\gamma j} \ .
\]
Thus, we obtain a theory of $M$ coupled SU$(2N)$ spin chains, described by
the Hamiltonian:
\begin{eqnarray}
H &=&
- \sum_x \sum_j (-1)^x \eta \ {\rm Tr} \left [ \Lambda S_j (x)
         \right ] \cr &&
+ \sum_x \sum_j \ J_j
       \ {\rm Tr} \left [ S_j (x+1) S_j (x) \right ] \cr &&
+ \sum_x \sum_{j < k} \ J_{jk}
       \ {\rm Tr} \left [S_j (x) S_k (x) \right ] \ ,
\label{spinH}
\end{eqnarray}
where $J_j = \langle |t_j|^2 \rangle$, $J_{jk} = - \langle
|t_{jk}|^2 \rangle$ and $\Lambda$ is a $2N \times 2N$ diagonal matrix
with $\Lambda^\alpha_\alpha = 1$ for $\alpha \le N$ and
$\Lambda^\alpha_\alpha = -1$ for $N < \alpha \le 2N$.  Also ${\rm Tr}
\left [ A B \right ] \equiv \sum_{\alpha \beta} A^\beta_\alpha
B^\alpha_\beta$ for any matrix $A$ and $B$.
Since $H$ commutes with the local density $n_j (x) = \sum_{\alpha}
\psi^{\dagger}_{\alpha j} \psi_{\alpha j}$ of each chain $j$, Hilbert
spaces corresponding to different $\{ n_j(x) \}$ decouple.
Consider first the case of decoupled chains, described by the first two terms
of the Hamiltonian of Eq.~(\ref{spinH})\cite{dhlee,affleck}.
In this case, the ground state of each chain belongs to the Hilbert space
where $n_j (x) = N$ for all $x$ \cite{dhlee}.
In this Hilbert space, a particular representation for SU$(2N)$ is
realized, characterized by a Young tableau with a single column of
length $N$ \cite{dhlee,affleck}.
In the above mapping from Eq.~(\ref{replicaS}) to Eq.~(\ref{spinH}),
we drop constant terms expecting that the ground state of the Hamiltonian
{}~(\ref{spinH}) still lies in the Hilbert space
where $n_j (x) = {\rm const.}$ for all $x$.

The above Hamiltonian gives nearest-neighbor antiferromagnetic
couplings ($J_j > 0$) in each spin chain $j$. On the other hand, for
each $x$, the coupling between spins on different chains, $j$ and $k$,
is ferromagnetic ($J_{jk} < 0$).
In terms of the original problem, the localization
corresponds to the existence of an
energy gap in the spectrum of this spin chain (in the replica limit of
$N \rightarrow 0$).


Before considering the limit of many coupled chains, it is worthwhile
to consider first the system for SU$(2)$ with two coupled chains only
($M = 2$ and $N = 1$). In this case, we have two antiferromagnetic
spin-${1 \over 2}$ chains with ferromagnetic coupling ($J_{12} = K_\perp < 0$)
between the chains (see Fig.3).
We can consider a larger parameter space where the
antiferromagnetic coupling along each spin chain alternates from bond
to bond between two values, $K_1$ and $K_2$. This means that each spin
chain corresponds to a single-channel network model with non-zero Hall
conductance. In order to observe the constraint of zero overall Hall
conductance for the present problem, we require the alternation of the
bond strengths on the two chains to be staggered. We are therefore
coupling together two single-channel networks with Hall conductances
of opposite signs. In other words, we have $J_1(x = {\rm even}) =
J_2(x = {\rm odd}) = K_1$ and $J_1 (x = {\rm odd}) = J_2 (x = {\rm
even})= K_2$ with $K_1,K_2>0$.  We can now study the system as a
function of the parameter $\Delta = (K_1 - K_2)/(K_1 + K_2)$. Since
the question of localization is believed to be independent of
microscopic details, it is interesting to ask whether the spectrum of
the two coupled chains has an energy gap irrespective of the
parameters $\Delta$ and $K_\perp$ (as long as the latter is finite).

As already mentioned, setting $K_\perp = 0$ results in two independent
single-channel networks. This has been studied in detail by D.H.~Lee
and coworkers \cite{wang,dhlee,dhlee2} in the context of the plateau
transition in quantum Hall systems. Each of these networks should
exhibit delocalization at $\Delta=0$. For finite $|\Delta|$, the
alternating coupling gives rise to dimerization, i.e., a
spin-Peierls phase. Instead of changing $\Delta$ to destroy the
gapless phase, we may also switch on the interchain ferromagnetic
coupling. This is a relevant perturbation to the decoupled gapless
spin chains. Therefore, in the case of $\Delta=0$, we expect the
low-energy physics of the coupled system to be described by the
opposite limit of strong coupling ($|K_\perp| \gg K_1=K_2$) where one
has a uniform antiferromagnetic spin-1 chain. It is known that this
system is always in a Haldane phase\cite{millis,watanabe}.

It remains to consider the case of finite $\Delta$ where we couple two
chains in spin-Peierls phases in such a way that their dimerizations
are staggered. This case has been studied numerically with
the conclusion that the coupling does not destroy the energy gap.
Instead, the system is believed to cross over from the dimerized phase
to a Haldane phase\cite{hida} as $K_\perp$ becomes much stronger than
$K_1$ and $K_2$. Thus, we can see that there are no phase transitions
along the boundaries of the $\Delta - K_\perp$ plane. This suggests
that there are also no phase boundaries within this $\Delta - K_\perp$
plane. In other words, the existence of an energy gap in this system
is indeed independent of these parameters of the mode, and it points
to the absence of a delocalization transition in the network model.

Let us return to the general case of arbitrary $M$ and $N$.
As in the SU$(2)$ case considered above, the interchain ferromagnetic
coupling between any two of $M$ spin chains is a relevant perturbation
in the renormalization-group sense. As the number $M$ of spin chains
becomes large, we expect the $M$ SU$(2N)$ spins at the same $x$-site to
form a totally symmetric representation of SU$(2N)$, which
corresponds to a Young tableau with $M$ columns of length $N$.
This strong-coupling limit can be regarded as a single SU$(2N)$ spin chain
where the spin at each site is in this large-$M$ representation. The
residual coupling between the spins at different sites remain
antiferromagnetic.


In the large-$M$ limit (as $\eta \rightarrow 0$), this
antiferromagnetic spin chain is equivalent to the following ${\rm U}(2N) /
{\rm U}(N) \times {\rm U}(N)$ sigma model \cite{dhlee,dhlee2,affleck}
\begin{equation}
{\cal L} = {M \over 16} {\rm Tr} (\partial_{\mu} Q)^2
+ {M \over 16} \epsilon_{\mu \nu} {\rm Tr} \left [
Q \partial_{\mu} Q \partial_{\nu} Q \right ] \ ,
\label{sigma}
\end{equation}
where $\mu = x,y$, $Q = u^{\dagger} \Lambda u$, and
$u$ is a $2N \times 2N$ unitary matrix.
The second term is a topological term\cite{pruisken,affleck}
and satisfies
\begin{equation}
\int d\tau dx \ \epsilon_{\mu \nu} {\rm Tr} \left [
Q \partial_{\mu} Q \partial_{\nu} Q \right ] =
16 \pi i m \ ,
\label{topology}
\end{equation}
where $m$ is an integer. Since $M$ is always an even integer in our
problem, the topological term is always irrelevant and can be dropped
in the Lagrangian.  Therefore, the system can be described by a
${\rm U}(2N) / {\rm U}(N) \times {\rm U}(N)$ sigma model ${\it without}$ a
topological
term.  Zhang and Arovas\cite{zhang} have argued that a term describing
logarithmic interactions between topological densities may exist and
lead to a gapless phase, similar to the Kosterlitz-Thouless
phase.
However, we have been unable to find any evidence for
such behavior in this spin-chain mapping.
In particular, for the $N=1$ problem of SU$(2)$
spins, it is known that the spin chain has a gap in its spectrum for
all even $M$ (integer spins).
Thus, one can see that we do not have the
long-range interaction between topologocal densities
in the case of $N = 1$.
According to Zhang and Arovas \cite{zhang}, there should be a term
describing the long-range interaction for arbitrary $N$.
Therefore, unless the theory is highly nonanalytic in terms of $N$
so that the replica limit is not well defined,
we expect the absence of
long-range interaction between topological densities even for
arbitrary $N$.

The beta function of the non-linear sigma model
(without a topological term) in the $N \rightarrow
0$ limit was evaluated by Hikami\cite{hikami} and is given by
\begin{equation}
\beta (1/M) \equiv {d \ {\rm ln} (1/M) \over d \ {\rm ln} L} =
2 (1/M)^2 + 6 (1/M)^4 + {\cal O} \biglb( (1/M)^6 \bigrb) \ ,
\label{betaM}
\end{equation}
where $L$ is the size of the system. Using the relation
Eq.~(\ref{conductance}), we can obtain the beta function for the
dimensionless conductance $g = G/(e^2/h)$:
\begin{equation}
\beta (g) \equiv {d \ {\rm ln} g \over d \ {\rm ln} L} =
- {2 \over (2 \pi g)^2} - {6 \over (2 \pi g)^4}
+ {\cal O} (g^{-6}) \ .
\label{betag}
\end{equation}
This is the known beta function for the dimensionless conductance in
the unitary ensemble\cite{aronov,hikami}.
We therefore believe that all the states are localized and
that the localization length in the large-conductance limit is given
by\cite{palee,comment}
\[
\xi \propto e^{\pi^2 g^2} \ .
\]

In summary, we have used a multi-channel network model to study the
localization properties of non-interacting fermions in a smoothly
varying random magnetic field with zero average. Using the replica
trick, the network with an even number of channels is mapped onto an
even number of coupled SU$(2N)$ spin chains in the $N
\rightarrow 0$ limit, where $N$ is the replica index. In the
large-conductance limit where the number of channels is large, this
system is equivalent to a particular representation of the ${\rm U}(2N) /
{\rm U}(N) \times {\rm U}(N)$ sigma model ($N\rightarrow 0$). Thus, we confirm
that the beta function $\beta (1/M)$ of this model is consistent with
the known $\beta (g)$ of the unitary ensemble. This result, which
suggests that all states are localized, agrees with perturbative
calculations\cite{aronov,falko} for random fields with short-ranged
correlations, and should be contrasted with the prediction of Zhang
and Arovas
\cite{zhang} that the system is critical for bare conductances greater
than $e^2/h$.

\acknowledgments
We would like to thank P.~A.~Lee and X.-G.~Wen for
helpful discussions.
We are supported by NSF grant No.~DMR-9411574 (Y.B.K.),
NSF grant No.~DMR-9216007 (A.F.), and
EPSRC/NATO Fellowship (D.K.K.L.).

\begin{figure}
\caption{
A schematic illustration of a network in the problem
of the random magnetic field. The two channels are
denoted by solid and dashed lines.}
\label{2channel}
\end{figure}

\begin{figure}
\caption{
A representation of the network in terms of chiral
fermions with alternating directions in the case of
$M = 2$.
The two channels are denoted by solid and
dashed lines.}
\label{fermion2channel}
\end{figure}

\begin{figure}
\caption{
A schematic representation of the two coupled spin
chains ($M = 2$). Note that $J_j (x) = J_j (x + 2)$
[$J_j (x) = \langle |t_j|^2 \rangle (x)$ and $j = 1,2$].
Also, $J_1 (x = {\rm even}) = J_2 (x = {\rm odd})
= K_1$ and $J_1 (x = {\rm odd}) = J_2 (x = {\rm even})
= K_2$. $J_{12} = \langle |t_{jk}|^2 \rangle = K_\perp$
is independent of $x$.}
\label{spin}
\end{figure}


\begin{references}
\bibitem{comment} It should be noted that we
are not concerned with evaluation of correlation functions
at a conformally invariant fixed point and so we
believe that we do not encounter the problems of replica limit
discussed by Affleck \cite{affleck} in relation to the
quantum Hall plateau transition.
\end{references}
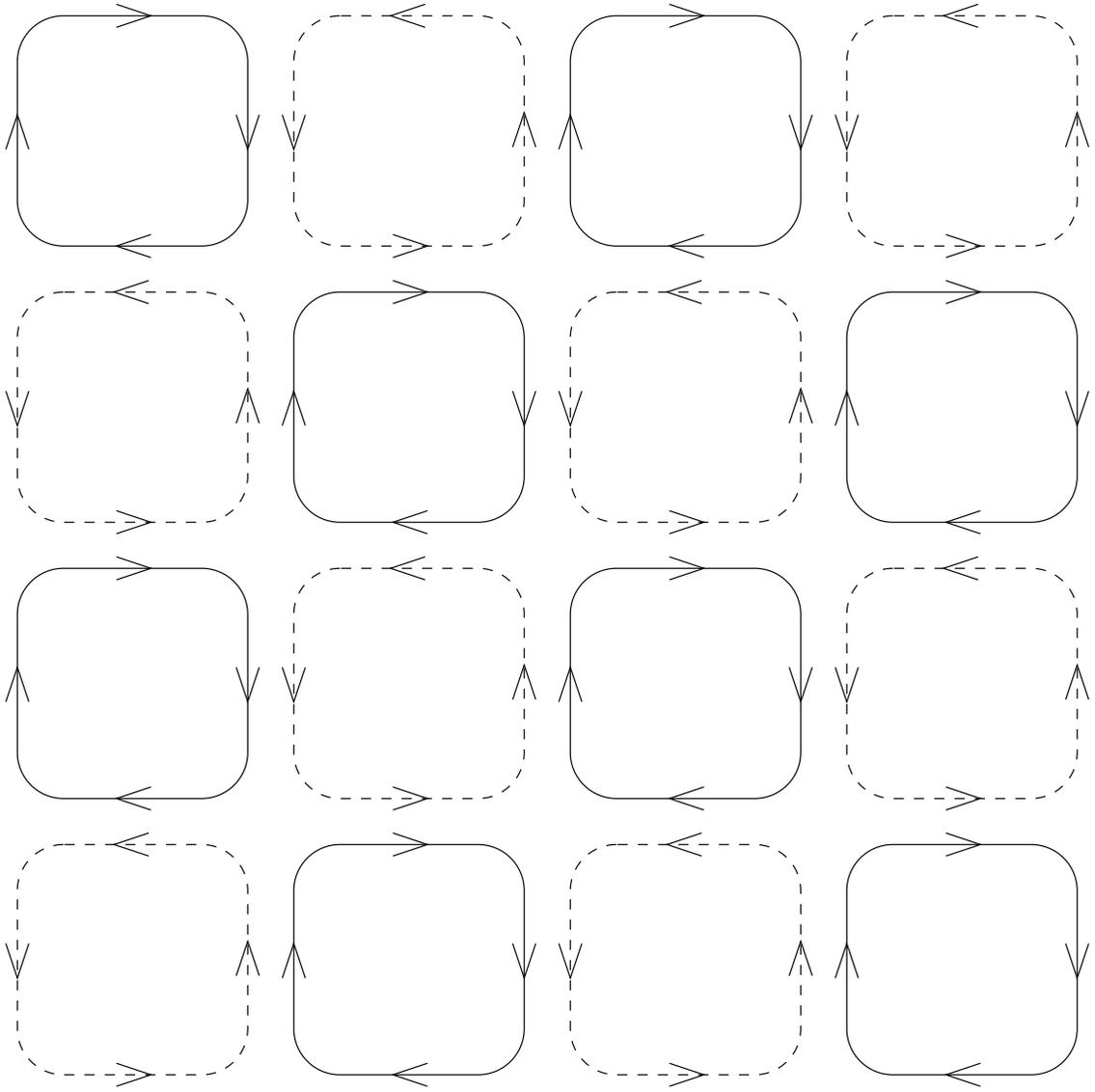
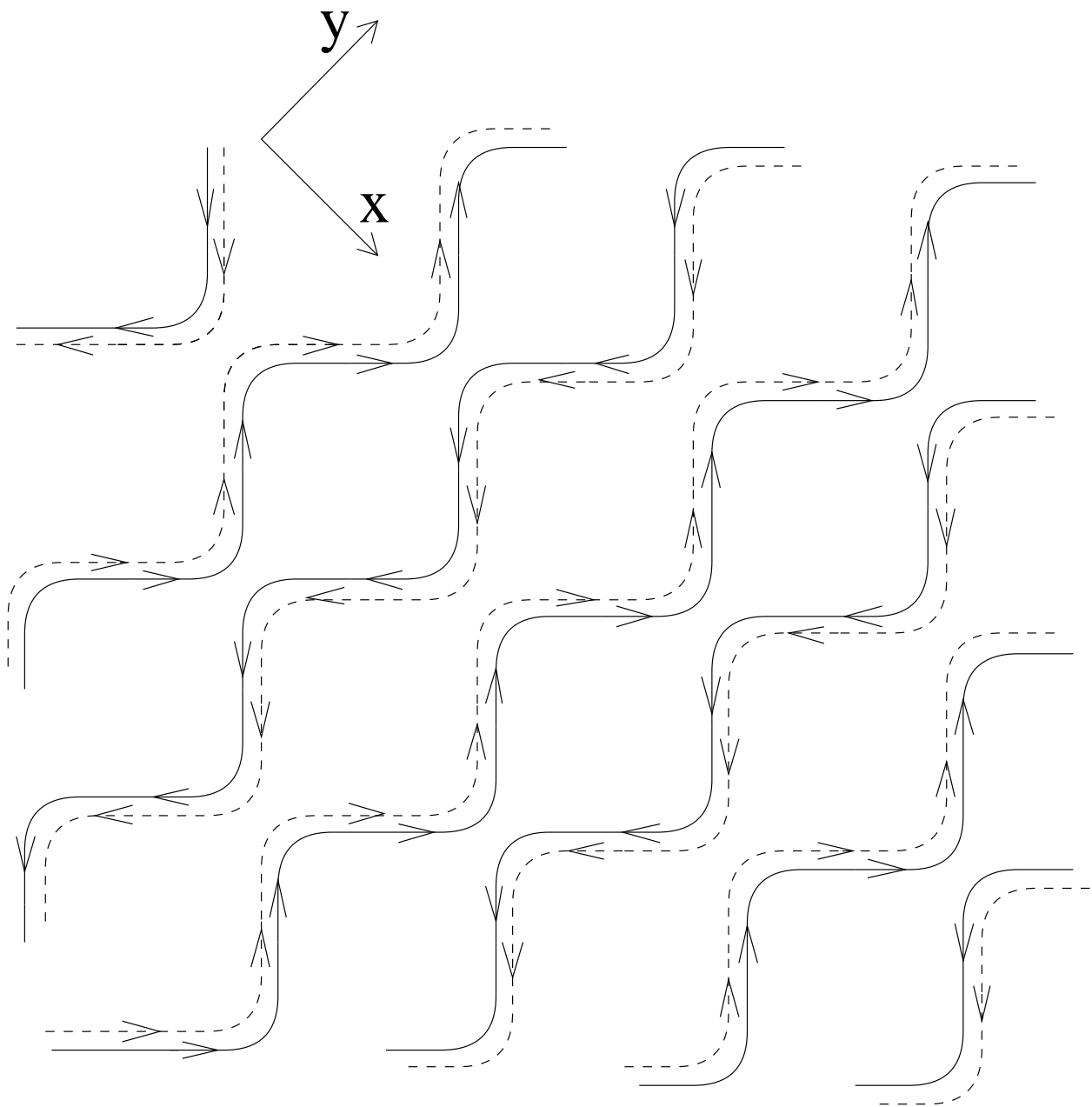
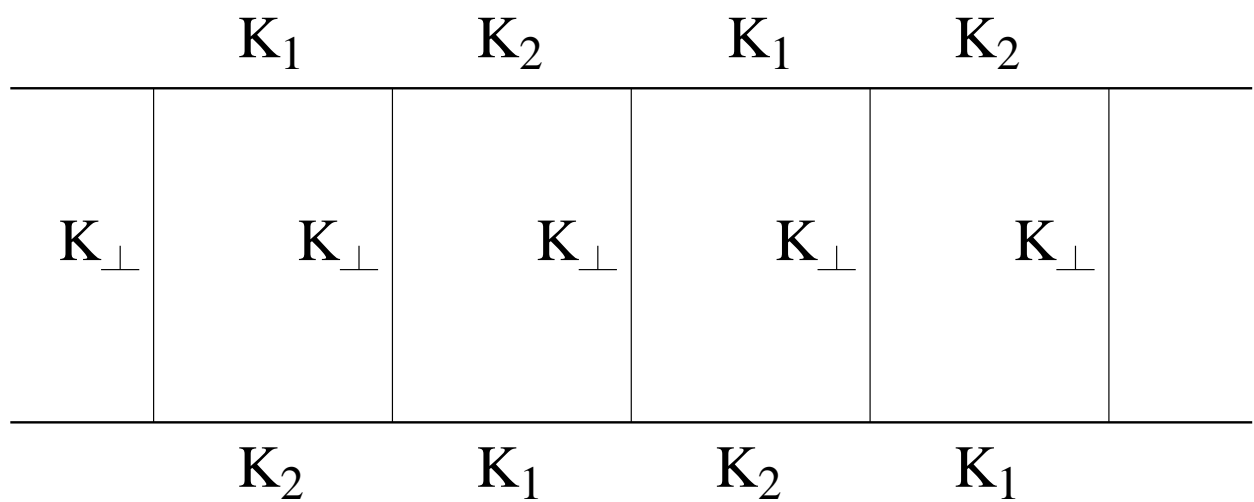
\end{document}